\documentclass[preprint,a4paper,12pt]{elsarticle}
\usepackage{axodraw}
\usepackage{slashed}
\usepackage{amssymb,amsmath,amsbsy}
\usepackage{mathrsfs}
\usepackage{bm}
\usepackage{hyperref}
\usepackage[top=1.2in, bottom=1.2in,left=0.9in,includefoot]{geometry}

\newcounter{bla}

\def\code{{\tt SUSY\_FLAVOR}}
\def\MMI{{\tt MassToMI}}
\def\webpage{{\tt www.fuw.edu.pl/masstomi}}

\def\eq#1{eq.~(\ref{#1})}

\def\eqs#1#2{eqs.~(\ref{#1}) and (\ref{#2})}

\def\eqst#1#2{eqs.~(\ref{#1})--(\ref{#2})}
\def\Eqst#1#2{Eqs.~(\ref{#1})--(\ref{#2})}

\def\Refs#1{refs.~\cite{#1}}

\newcommand{\bea}{\begin{eqnarray}}
\newcommand{\eea}{\end{eqnarray}}

\usepackage{color}

\definecolor{orange}{rgb}{0.9,0.2,0}
\definecolor{brown}{rgb}{0.7,0.3,0.2}
\definecolor{fuxia}{rgb}{1,0,1}
\definecolor{skyblue}{rgb}{0,0.1,0.9}
\definecolor{violetred}{rgb}{0.8,0.13,0.56}
\definecolor{deeppink}{rgb}{1.00,0.08,0.5}
\definecolor{pink}{rgb}{1.00,0.75,0.80}
\definecolor{orchid}{rgb}{0.85,0.44,0.84}
\definecolor{lightpink}{rgb}{1.00,0.71,0.76}
\definecolor{bluish}{rgb}{0,0.6,0.8}

\journal{Computer Physics Communications}

\begin{document}

\begin{frontmatter}

\title{MassToMI - a Mathematica package for an automatic Mass
  Insertion expansion}

\author[a]{Janusz Rosiek\corref{author}}

\cortext[author] {Corresponding author. \textit{E-mail address:}
  janusz.rosiek@fuw.edu.pl}

\address[a]{Institute of Theoretical Physics, Physics Department,
  Warsaw University, Pasteura 5, 02-093 Warsaw, Poland}

\begin{abstract}
We present a Mathematica package designed to automatize the expansion
of transition amplitudes calculated in the mass eigenstates basis
(i.e. expressed in terms of physical masses and mixing matrices) into
series of ``mass insertions'', defined as off-diagonal entries of mass
matrices in Lagrangian before diagonalization and identification of
the physical states.  The algorithm implemented in this package is
based on the general ``Flavor Expansion Theorem'' proven in
Ref.~\cite{FET}. The supplied routines are able to automatically
analyze the structure of the amplitude, identify the parts which could
be expanded and expand them to any required order.  They are capable
of dealing with amplitudes depending on both scalar or vector
(Hermitian) and Dirac or Majorana fermion (complex) mass matrices. The
package can be downloaded from the address \webpage.
\end{abstract}


\end{frontmatter}

\newpage

\noindent {\bf PROGRAM SUMMARY}

\begin{small}
\noindent
{\em Manuscript Title:} \\ 
MassToMI - a Mathematica package for an automatic Mass Insertion
expansion \\
{\em Authors:} Janusz Rosiek \\
{\em Program Title:} MassToMI v1.0 \\
{\em Journal Reference:} \\
{\em Catalogue identifier:} \\
{\em Licensing provisions:} None \\
{\em Programming language:} Mathematica 10.2 (earlier versions should
work as well) \\
{\em Computer:} any running Mathematica \\
{\em Operating system:} any running Mathematica \\
{\em RAM:} allocated dynamically by Mathematica, at least 4GB total
RAM suggested \\ 
{\em Number of processors used:} allocated dynamically by Mathematica \\
{\em Supplementary material:} None 
{\em Classification:}\\ \begin{tabular}{ll}
11.1 & General, High Energy Physics and Computing,\\
5 & Computer Algebra. \\
\end{tabular}\\
{\em External routines/libraries:} Wolfram Mathematica program \\
{\em Subprograms used:} None \\
{\em Nature of problem:}\\ 
Automatized expansion of QFT transition amplitude calculated in mass
eigenstates basis into power series of off-diagonal elements of mass
matrices of the interaction basis Lagrangian. \\
{\em Solution method:}\\ 
Implementation (as the Mathematica package) of the algebraic algorithm
``Flavor Expansion Theorem'', formulated and proven in Ref.~[1] given
below.  \\
{\em Restrictions:} None\\
{\em Unusual features:} None \\
{\em Additional comments:} None \\
{\em Running time:} depending on complexity of the analyzed
expression, from seconds for simple problems to hours for complicated
amplitudes expanded to high order (using Mathematica 10.2 running on a
personal computer) \\

\end{small}

\newpage

\section{Introduction}
\setcounter{equation}{0}
\label{sec:intro}

Quantum Field Theory (QFT) models are usually defined by specifying
the Lagrangian of the theory. Such definition is not unique, in the
sense that it allows for transformations of the field basis, leading
to equivalent descriptions of the model, with different degrees of
freedom.  The calculation of transition amplitudes can be performed in
any basis, however two special choices are most convenient from the
practical point of view.

In many cases the Lagrangian of the model is initially constructed in
terms of fields having definite charges under some symmetry groups -
we call it {\em symmetry} or {\em interaction basis}.  However, in
general, fields defined in such a way do not correspond to physical
degrees of freedom of the theory, and some redefinitions have to be
performed in order to identify the physical fields (with the
spontaneous symmetry breaking being the typical example).

Another possibility is to use {\sl mass eigenbasis}, in which bilinear
terms (kinetic and mass matrices) in the Lagrangian have been
diagonalized and states of the theory correspond to physical particles
with definite mass. The transformation from the initial basis to the
mass eigenstates basis is performed by unitary rotations (``mixing
matrices'') in the field space.  Perturbative calculations of the
amplitudes in the mass eigenstates basis lead to results expressed in
terms of physically measurable quantities, i.e. physical masses and
the elements of the mixing matrices.  They are usually more compact as
compared to those obtained in any other basis and best suited for
numerical computations. However, the analytical dependence of such
amplitudes on the initial interaction basis is typically complicated
and difficult to use for qualitative interpretation.

For the latter purpose, it is often useful to have analytic
expressions for the transition amplitudes calculated directly in the
interaction basis. They can be obtained using two different methods.
Firstly, one can perform an independent diagrammatic calculation of
the amplitude using the approximation commonly referred to as the {\sl
  Mass Insertion Approximation} (MIA)~\cite{Hagelin:1992tc,
  Gabbiani:1996hi, Misiak:1997ei}.  In this approach, diagonal
elements of the mass matrices are absorbed into the definition of
(unphysical) massive propagators and the amplitude is, at every loop
order, expanded in an infinite series of non-diagonal elements of mass
matrices, commonly referred to as {\sl mass insertions} (MI).
Alternatively, as proven in Ref.~\cite{FET}, the MIA result can be
obtained directly from the mass eigenstates amplitude employing the
purely algebraic technique coined in Ref~\cite{FET} as the ``Flavor
Expansion Theorem'' (FET)\footnote{The first non-trivial order of the
  FET expansion, with applications in MSSM flavor physics, have been
  presented in~\Refs{Buras:1997ij}.  Higher orders could be also
  obtained using the standard quantum mechanic perturbation theory,
  applied to mass matrix eigenstates problem (see
  e.g.~\cite{Crivellin:2010gw}), but in this case it is very difficult
  to get the simple closed expressions.  }.  The last method, the FET
expansion, has two important advantages.  Firstly, it allows to avoid
the Feynman diagram calculation with mass insertions, which is usually
tedious and prone to errors or omissions of important terms. Secondly,
it can become, to large extent, automatized.

In this paper we describe the \MMI{} package, written with the use of
Mathematica~\cite{mathematica102} symbolic manipulation language and
designed to perform automatically the MI expansion of QFT amplitudes
evaluated in mass eigenstates basis, provided that they are coded
using a specific format which could be parsed by the \MMI{} routines.
The results are given in terms of MI powers and the so-called {\sl
  divided differences}~\cite{de2005divided} of the loop functions.
The package is able to expand any type of amplitude for Hermitian
(scalar or vector) or general complex (fermion) mass matrices, to any
requested MI order.

The paper is organized as follows. In Sec.~\ref{sec:basis} we
formulate the algorithm used for the expansion. In
Sec.~\ref{sec:syntax} we present the syntax for defining the
amplitudes in the \MMI{} package, the routines provided for the users
and the output format.  In Section~\ref{sec:example} we illustrate the
applications of the \MMI{} package with several examples, finally we
conclude in Sec.~\ref{sec:summary}.  The \MMI{} Mathematica code can
be downloaded from the address
\begin{center}
\webpage
\end{center}

%

\section{Flavor and mass amplitudes in QFT calculations}
\setcounter{equation}{0}
\label{sec:basis}

\subsection{Mass matrices in physical QFT models}
\label{sec:mass}

The physical spectrum of any QFT model is defined by the structure of
the quadratic terms in the Lagrangian. Assuming that quadratic kinetic
(momentum-dependent) terms have been transformed to the canonical form
by appropriate field and coupling redefinitions and, if necessary, the
Spontaneous Symmetry Breaking (SSB) mechanism has been used to
identify physical degrees of freedom, three type of mass matrices can
appear in the Lagrangian:
\begin{enumerate}
\item Hermitian (squared) mass matrices for scalar and vectors fields,
  ${\bf M_S^2 = (M_S^2)^\dagger}$, diagonalized by a unitary
  transformation $\mathbf{Z}$:
\bea
{\bf Z^\dagger M_S^2 Z} = \ \mathbf{m_S^2} \ = \mathbf{diag}
(m_1^2,\ldots, m_n^2)\,.
\eea
\item General complex mass matrices for Dirac fermions ${\bf M_D}$,
  diagonalized by two unitary transformations $\mathbf{U,V}$:
\bea
\mathbf{V^\dagger\, M_D\, U}\ = \ \mathbf{m_D} \ = \ \mathbf{diag}
(m_1,\ldots, m_n)\,.
\eea
The matrices $\mathbf{V}$ and $\mathbf{U}$ diagonalize also the
Hermitian matrices ${\bf M_D M_D^\dag}$ and ${\bf M_D^\dag M_D}$,
through the transformations
\bea
{\bf V^\dagger\, M_D\: M_D^\dagger\, V} \ = \ {\bf U^\dagger\,
  M_D^\dagger\: M_D\, U} \ = \ {\bf m_D^2}\,.
\label{eq:uvdef}
\eea
\item Symmetric complex mass matrices for Majorana fermions, ${\bf M_N
  = M_N^T}$. In such case one can assume $\mathbf{U=V^\star=O}$
  in~\eq{eq:uvdef}, so that the mass matrix is diagonalized by a
  single unitary transformation $\mathbf{O}$:
\bea
\mathbf{O^T\, M_N\, O}\ = \ \mathbf{m_N} \ = \ \mathbf{diag}
(m_1,\ldots, m_n)\,.
\eea
The matrix $\mathbf{O}$ diagonalizes also the Hermitian matrix ${\bf
  M_N^\dag M_N}$,
\bea
{\bf O^\dagger\, M_N^\dagger\: M_N\, O} \ = \ {\bf m_N^2}\,.
\label{eq:odef}
\eea
\end{enumerate}
Note that the squared mass matrices of physical particles, i.e. ${\bf
  M_S^2, M_D\: M_D^\dagger, M_D^\dagger\: M_D}$ and ${\bf M_N^\dag\:
  M_N}$ must be (semi-) positive-definite for well-defined QFT
theories.

\subsection{Structure of the transition amplitudes}
\label{sec:ampl}

Applying the rotations ${\bf Z,U,V,O}$ to the field multiplets one
gets the Lagrangian of the theory in the basis of physical (mass
eigenstates) fields.  The tree level vertices and Feynman rules in
such basis depend on the elements of mixing matrices and on the
physical particle masses.  Consequently, all transition amplitudes (to
any loop order) are linear combinations of products of mixing matrices
and propagators, or loop integrals being the functions of physical
particle masses.  The dependence of such amplitudes on the initial
Lagrangian parameters is very complicated and non-linear. Thus, as
mentioned in the introduction, although they are compact in form and
better suited for numerical computations, it is often practical to use
the Mass Insertion expansion to recover an approximate, but simpler
direct dependence on the interaction basis Lagrangian parameters.

Such an expansion can be done with the use of Flavor Expansion
Theorem, formulated and proven in Ref.~\cite{FET}.  As argued
in~\cite{FET}, the mixing matrices of the internal particles of
Feynman diagrams can appear in amplitudes only in some specific
combinations, namely
\bea
\mathrm{Scalars, vector~bosons:}  && Z_{Bi}\, f(m_i^2) \, Z_{Ai}^{\star} \;,
\nonumber\\[1mm]
\mathrm{Dirac~fermions:}  && U_{Bi}\, f(m_i^2) \, U_{Ai}^{\star}, \, \,
V_{Bi}\, f(m_i^2) \, V_{Ai}^{\star},\;,\nonumber\\[1mm]&& U_{Bi}\, m_i f(m_i^2) \,
V_{Ai}^{\star},\,\, V_{Bi}\, m_i f(m_i^2) \,
U_{Ai}^{\star} \;,\nonumber\\[1mm]
\mathrm{Majorana~fermions:}  && O_{Bi}\, f(m_i^2) \, O_{Ai}^{\star}, \,\,
O_{Bi}\, m_i f(m_i^2) \, O_{Ai}\;,
\label{eq:mixtype}
\eea
where $f(m_i^2)$ represents symbolically the dependence of the
amplitude on the physical masses, at tree or loop level. In the
expression above, and similarly for other equations in the rest of the
paper, we assume a summation over the repeating indices (even if they
appear more than twice).

Assuming that the function $f$ is analytical, the combinations listed
above can be formally expressed as matrix elements being functions of
the squared mass matrices:
\bea
\mathrm{Scalars,~vector~bosons:}&&\nonumber\\
Z_{Bi}\, f(m_i^2) \, Z_{Ai}^{\star} &=&
f(\mathbf{M_S^2})_{BA}\;,
\label{eq:scalarFET}
\\[3mm]
\mathrm{Dirac~fermions:}&&\nonumber\\
U_{Bi}\, f(m_i^2) \, U_{Ai}^{\star} &=& f(\mathbf{M_D^\dag
  M_D})_{BA}\;,\nonumber\\[1mm]
V_{Bi}\, f(m_i^2) \, V_{Ai}^{\star} &=& f(\mathbf{ M_D M_D^\dag
})_{BA}\;,\nonumber\\[1mm]
V_{Bi}\, m_i f(m_i^2) \, U_{Ai}^{\star} &=& \left(\mathbf{M_D} \,
  f(\mathbf{M_D^\dag M_D})\right)_{BA} = \left( f(\mathbf{M_D
  M_D^\dag}) \, \mathbf{M_D} \right)_{BA}
\;,\nonumber\\[1mm]
U_{Bi}\, m_i f(m_i^2) \, V_{Ai}^{\star} &=& \left(\mathbf{M_D^\dag} \,
  f(\mathbf{M_D M_D^\dag})\right)_{BA} = \left( f(\mathbf{M_D^\dag
  M_D}) \, \mathbf{M_D^\dag} \right)_{BA}
\;,
\label{eq:fermionFET}
\eea
%
%
\bea
\mathrm{Majorana~fermions:}&&\nonumber\\
O_{Bi}\, f(m_i^2) \, O_{Ai}^{\star} &=& f(\mathbf{M_N^\dag M_N})_{BA} =
  f(\mathbf{M_N M_N^\dag})_{AB} \;,\nonumber\\[1mm]
O_{Bi}\, m_i f(m_i^2) \, O_{Ai} &=& \left(\mathbf{M_N^\dag} \,
f(\mathbf{M_N M_N^\dag}) \right)_{BA} = \left( f(\mathbf{M_N^\dag
  M_N}) \, \mathbf{M_N^\dag} \right)_{BA} \;,\nonumber\\[1mm]
O_{Bi}^\star\, m_i f(m_i^2) \, O_{Ai}^\star &=& \left(\mathbf{M_N} \,
f(\mathbf{M_N^\dag M_N}) \right)_{BA} = \left( f(\mathbf{M_N
  M_N^\dag}) \, \mathbf{M_N} \right)_{BA} \;.
\label{eq:majoranaFET}
\eea
RHS of all of the expressions above depends on the matrix elements of
functions of the hermitian matrices.  As proven in Ref.~\cite{FET},
such elements can be expanded in terms of ``divided differences''
$f^{[k]}$ of the function $f$, defined recursively as
\bea
f^{[0]}(x)&&\equiv f(x)\: \;,\nonumber\\[1mm]
f^{[1]}(x_0,x_1)&&\equiv\frac{f(x_0)-f(x_1)}{x_0-x_1}\: \;, \nonumber\\[1mm]
\ldots\nonumber\\
f^{[k+1]}(x_0,\dots ,x_k,x_{k+1})&&\equiv\frac{f^{[k]}(x_0,\dots ,
  ,x_{k-1},x_k)-f^{[k]}(x_0,\dots ,x_{k-1},x_{k+1})}{x_k-x_{k+1}} \;
\label{dddef}
\eea
or, for degenerate case (and analytical $f$),
\bea
\lim_{\{ x_0,\dots ,x_m\} \to \{ \xi,\dots,\xi \}}
f^{[k]}(x_0,\dots,x_k)=\frac{1}{m!}\frac{\partial^{m} }{\partial
  \xi^{m}}f^{[k-m]}(\xi,x_{m+1}\dots,x_k)\;.
\label{ddlimit}
\eea
The expansion can be done by splitting the mass matrices into diagonal
parts and off-diagonal mass insertions,
\bea
\mathbf{M^2}\,=\,\mathbf{M^2_0}\,+\,\mathbf{{\hat M}^2}\, ,\label{Ageneral}
\eea
where, by definition,
\bea
\begin{array}{l}
(M^2_0)_I\,\equiv M^2_{II}\;,\\[3mm]
{\hat M}^2_{IJ}\,\equiv\,M^2_{IJ},\qquad{\hat M}^2_{II}=0\;, \hspace{1cm}
(I,J=1,\dots,n)\;.
\end{array}\label{Adiag}
\eea
Then, for any analytical Hermitian matrix function $f({\bf {M^2}})$
(see~\cite{FET} for detailed assumptions), a matrix element will be
given by the expansion
\bea
f({\bf M^2})_{IJ} &=& \delta_{IJ}f((M^2_0)_I)\ +
f^{[1]}((M^2_0)_I,(M^2_0)_J)\:{\hat M}^2_{IJ}\nonumber\\[1mm]
&+& \sum_{K_1} f^{[2]}((M^2_0)_I,(M^2_0)_J,(M^2_0)_{K_1})\:
{\hat M}^2_{I{K_1}}{\hat M}^2_{{K_1}J}
\label{theorem}
\\[1mm]
& +& \sum_{K_1,K_2}
f^{[3]}((M^2_0)_I,(M^2_0)_J,(M^2_0)_{K_1},(M^2_0)_{K_2}) \: {\hat
  M}^2_{I{K_1}}{\hat M}^2_{{K_1}{K_2}}{\hat M}^2_{{K_2}J} + \ldots
\;,\nonumber
\eea
Notice that due to the definition~(\ref{Adiag}) terms proportional to
$\hat{M}^2_{II}$ will always vanish in the summation.

\subsection{Ambiguity in expansion of fermionic amplitudes}
\label{sec:ambi}

It is important to note that although the alternative forms of the RHS
in the last 2 lines of~\eqs{eq:fermionFET}{eq:majoranaFET}, appearing
in expansion of fermionic amplitudes, are order-by-order equivalent
when the function $f$ is expanded in a Taylor series, they are {\em
  not} order-by-order equivalent when FET expansion of~\eq{theorem}
for $f$ is used.  For example, lets consider the lowest order
expansion of the two forms of the RHS in the last line
of~\eq{eq:majoranaFET}.  It would lead to (for Majorana fermions
$\mathbf{M_N} = \mathbf{M_N^T}$):
\bea
&O_{Bi}^\star\, m_i f(m_i^2) \, O_{Ai}^\star = \left(\mathbf{M_N} \,
f(\mathbf{M_N^\dag M_N}) \right)_{BA} \to \left( \mathbf{M_N}
\right)_{BA} \, f\left(\left( \mathbf{M_N^\dag M_N}
\right)_{AA}\right) \nonumber\\
\label{eq:msel}
{\rm or}& \\
& O_{Bi}^\star\, m_i f(m_i^2) \, O_{Ai}^\star = \left( f(\mathbf{M_N
  M_N^\dag}) \, \mathbf{M_N} \right)_{BA} \to ( \mathbf{M_N} )_{BA} \,
f\left(\left(\mathbf{M_N M_N^\dag}\right)_{BB}\right)\nonumber
\eea
with both forms obviously different for $A\ne B$.

\begin{figure}[htbp]
\begin{center}
\begin{picture}(150,60)(0,-15)
\ArrowLine(0,0)(30,0)
\ArrowLine(30,0)(90,0)
\ArrowLine(90,0)(120,0)
\DashArrowArc(60,0)(30,0,180){4}
\Vertex(30,0){2}
\Vertex(90,0){2}
\Text(60,-10)[c]{$\chi_i$}
\Text(60,40)[c]{$\tilde\nu_K$}
\Text(-10,0)[c]{$l_I$}
\Text(130,0)[c]{$l_J$}

\end{picture}
\caption{Sneutrino-chargino loop contributing to the lepton
  self-energy in the MSSM. $I,J$ denote external lepton
  flavors. \label{fig:cse}}
\end{center}
\end{figure}
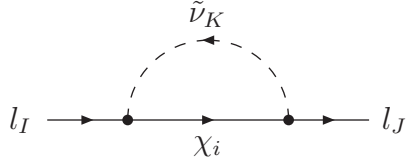

The correct choice of the form of the RHS
of~\eqs{eq:fermionFET}{eq:majoranaFET} and minimal order of expansion
reproducing the exact result with sufficient accuracy depends on a
given amplitude. To illustrate it with a realistic example, lets
consider the chargino contribution to the lepton self energy in the
Minimal Supersymmetric Standard Model (MSSM), illustrated in
Fig.~\ref{fig:cse}.

For the purpose of our example lets consider only the scalar part of
the lepton self-energy, $\Sigma_S^{IJ} = \Sigma^{IJ}_{SL} P_L +
\Sigma^{JI*}_{SL} P_R$, where, neglecting the small external momenta,
\bea
\Sigma_{SL}^{IJ} = g_2 Y_l^I \sum_{K=1}^3 \sum_{i=1}^2 Z_v^{IK}
Z_v^{JK*} V_C^{2i*} U_C^{1i} m_{\chi_i} b_0(m_{\tilde\nu_K}^2,
m_{\chi_i}^2) \equiv g_2 Y_l^I \sum_{K=1}^3 Z_v^{IK} Z_v^{JK*} F_K
\label{eq:cse}
\eea
where $b_0$ is the 2-point loop function, by $Y_l$ we denote the
lepton Yukawa coupling, $Z_v$ is the unitary matrix diagonalizing the
sneutrino mass matrix and $V_C,U_C$ diagonalize the chargino mass
matrix (for details of notation see Refs.~\cite{Rosiek:1989rs,
  Rosiek:1995kg}):
\bea
V_C^\dag
\left(
\begin{array}{cc}
M_2 & \frac{g_2 v_2}{\sqrt{2}} \\
\frac{g_2 v_1}{\sqrt{2}} & \mu \\
\end{array}
\right) U_C \equiv  V_C^\dag M_C U_C = {\rm diag}(m_{\chi_1}, m_{\chi_2})\;.
\eea
In addition by $F_K$ we denoted the factors in~\eq{eq:cse} depending
only on chargino mixing matrices:
\bea
F_K = \sum_{i=1}^2  V_C^{2i*} U_C^{1i} m_{\chi_i} b_0(m_{\tilde\nu_K}^2,
m_{\chi_i}^2) \; .
\label{eq:fk}
\eea
Expanding $F_K$ to the first MI order, one gets two possible
expressions, corresponding to two forms of the RHS
of~\eq{eq:fermionFET}. They read either as
\bea
F_K^{(1)} &\approx& M_C^{21*} b_0\left(m_{\tilde\nu_K}^2,(M_C
M_C^\dag)^{22}\right) + M_C^{11*} (M_C M_C^\dag)^{12}\,
b_0^{[1]}\left(m_{\tilde\nu_K}^2,(M_C M_C^\dag)^{11}, (M_C
M_C^\dag)^{22}\right)\nonumber\\[3mm]
& = & \frac{g_2v_1}{\sqrt{2}}\left[ b_0\left(m_{\tilde\nu_K}^2,
  |\mu|^2 + \frac{g_2^2 v_1^2}{2} \right) \right. \nonumber\\
& +& \left.  \left( |M_2|^2 + \frac{v_2}{v_1} \mu^* M_2^*\right)
  c_0\left(m_{\tilde\nu_K}^2, |M_2|^2 + \frac{g_2^2 v_2^2}{2}, |\mu|^2
  + \frac{g_2^2 v_1^2}{2} \right) \right]
\label{eq:fk1}
\eea
or as
\bea
F_K^{(2)} &\approx& M_C^{21*} b_0\left(m_{\tilde\nu_K}^2,(M_C^\dag
M_C)^{11}\right) + M_C^{22*} (M_C^\dag M_C)^{12} \,
b_0^{[1]}\left(m_{\tilde\nu_K}^2,(M_C^\dag M_C)^{11}, (M_C^\dag
M_C)^{22}\right)\nonumber\\[3mm]
& = & \frac{g_2v_1}{\sqrt{2}}\left[ b_0\left(m_{\tilde\nu_K}^2,
  |M_2|^2 + \frac{g_2^2 v_1^2}{2} \right) \right. \nonumber\\
&+& \left.  \left( |\mu|^2 + \frac{v_2}{v_1} \mu^* M_2^*\right)
  c_0\left(m_{\tilde\nu_K}^2, |M_2|^2 + \frac{g_2^2 v_1^2}{2}, |\mu|^2
  + \frac{g_2^2 v_2^2}{2} \right) \right]
\label{eq:fk2}
\eea
where we used the relation connecting divided difference of the
2-point loop function with the 3-point loop function,
$b_0^{[1]}(m_1^2,m_2^2,m_3^2) = c_0(m_1^2,m_2^2,m_3^2)$.

In the lowest MI order arguments of $b_0$ function
in~\eqs{eq:fk1}{eq:fk2} differ, which could be numerically important
if $\mu$ and $M_2$ are splitted significantly comparing to some
average scale $M_{SUSY}$.  However, using the identity (holding for
vanishing external loop momenta):
\bea
b_0(m_1^2,m_2^2) + m_3^2\, c_0(m_1^2,m_2^2,m_3^2) = b_0(m_1^2,m_3^2) +
m_2^2 \, c_0(m_1^2,m_2^2,m_3^2)\;,
\eea
one can see that in the 1st MI order~\eqs{eq:fk1}{eq:fk2} differ only
by terms of the order of $g_2^2v_{1(2)}^2/M_{SUSY}^2 \sim
M_W^2/M_{SUSY}^2$ (so of the higher order in MI expansion), in
practice usually negligible in calculations of the leptonic
transitions in the MSSM.

\section{Expanding amplitudes with \MMI{} package}
\setcounter{equation}{0}
\label{sec:syntax}

\Eqst{eq:scalarFET}{eq:majoranaFET} and~(\ref{theorem}) allow to
expand any transition amplitude calculated in the mass eigenstates
basis as a series in mass insertions powers.  However, for realistic
models with many fields it may lead to lengthy expressions, requiring
tedious and error-prone calculations.  To facilitate the problem, the
\MMI{} package automatizes such calculations.

\subsection{\MMI{} installation}

The \MMI{} package does not require any special installation
procedures.  It should be unpacked to the directory accessible to
Mathematica, or installed system-wide using {\em Install} command from
Mathematica menu.  Then the package can be loaded using the
commands\\[2mm]
{\tt Needs["MassToMI`"];}\\[1mm]
or\\[1mm]
{\tt << MassToMI.m}

\subsection{Syntax for constructing amplitudes in \MMI{}}

In general, any amplitude, 1PI irreducible or reducible, can be
constructed as a sum of products of mixing matrices of scalars,
fermions or vector bosons, physical particle masses, loop integrals
and other factors which are not affected by the FET expansion.  To
identify the combinations of objects relevant for the expansion
procedure, \MMI{} package require them to be denoted using the special
syntax, listed in Table~\ref{tab:syntax} (other symbolic or numerical
factors in the amplitude, not listed in Table~\ref{tab:syntax}, are
treated by program as constants).

\begin{table}[htbp]
\begin{tabular}{ll}
\hline
Object and its syntax in \MMI & Arguments \& examples\\
\hline\hline
Scalar (vector) mixing matrix $Z$ \\
~~ {\tt SMIX[P,i,j]} & $P$ - particle symbol \\
& $i,j$ - particle indices\\[2mm]
~~ Example: & $Z_D^{Ij} =$ {\tt SMIX[D,I,j]} \\
& $Z_U^{Ij*} =$ {\tt Conjugate[SMIX[U,I,j]] }\\[3mm]
\hline
Left Dirac fermion mixing matrix $V$ \\
~~  {\tt FMIXL[P,i,j]} & $P$ - particle symbol \\
& $i,j$ - particle indices\\[2mm]
~~ Example: & $V_D^{Ij} =$ {\tt FMIXL[D,I,j]} \\[3mm]
\hline
Right Dirac fermion mixing matrix $U$ \\
~~  {\tt FMIXR[P,i,j]} & $P$ - particle symbol \\
& $i,j$ - particle indices\\[2mm]
~~Example: & $U_Q^{Ij*} =$ {\tt Conjugate[FMIXR[Q,I,j]]} \\[3mm]
\hline
Majorana fermion mixing matrix $O$ \\
~~  {\tt NMIX[P,i,j]} & $P$ - particle symbol \\
& $i,j$ - particle indices\\[2mm]
~~Example: & $O_Q^{Ij} =$ {\tt NMIX[Q,I,j]}\\[3mm]
\hline
physical particle mass  \\
~~  {\tt MASS[P,i] } & $P$ - particle symbol \\
& $i$ - particle index \\[2mm]
~~Example: & $m_Q^I =$ {\tt MASS[Q,I]}  \\[3mm]
\hline
general loop function \\
~~ {\tt LOOP[name,\{\{P1,i1\},\ldots\},\{a,\ldots \}] } & $name$ -
name of function \\
& $ \{ \{P1,i1\},\ldots \} $ - physical loop masses \\
& given as the list of (particle,index) pairs\\
& $ \{a,\ldots \} $ - optional other arguments \\[2mm]
~~Examples: & $b_{21}(p,m_{U_i}^2,m_{D_j}^2) =$ \\
& {\tt LOOP[b21,\{\{U,i\},\{D,j\}\},\{p\}]} \\[2mm]
 & $c_{0}(p,q,m_{A_i}^2,m_{B_j}^2,m_{C_k}^2) =$ \\
& {\tt LOOP[c0,\{\{A,i\},\{B,j\},\{C,k\}\},\{p,q\}]} \\[3mm]
\hline
\end{tabular}
\caption{Syntax of objects used to constructing amplitudes in \MMI{}
  package~\label{tab:syntax}}
\end{table}

The sub-expressions in the amplitude which could be expanded in terms
of MIs need to have one of the forms listed on the LHS
of~\eqst{eq:scalarFET}{eq:majoranaFET}. Some further remarks are in
order here:
\begin{itemize}
\item As in the analytical expressions
  of~\eqst{eq:scalarFET}{eq:majoranaFET}, it is assumed that indices
  of particles appearing more than once in the amplitude, written as
  Mathematica code, are summed over without the need of specifying the
  sum explicitly.
\item The amplitude does not need to be 1PI, i.e. products of loop
  functions are allowed.
\item Only linear fermion mass powers (or multi-linear for amplitudes
  involving more than one fermion) can appear as factors in the
  amplitude, all even powers of masses should be included in the
  definitions of the loop functions.
\item Eq.~(\ref{dddef}) defines the divided differences for functions
  of one variable. The generalization to many variables is obvious:
  one should calculate the divided differences of the required order
  with respect to each variable separately.  However, complications
  arise when some of the particles in the loop have equal masses (as
  for example in the case of flavor-diagonal photon couplings) and
  several arguments of loop function are identical. Then, the
  generalization of eq.~\ref{dddef} is less trivial and leads to some
  form of ``Leibniz-like'' rule, which combinatorial complication
  grows quickly with the number of identical arguments and the order
  of divided difference. \MMI{} package does not allow for repeating
  arguments of the loop functions, assuming if necessary they have
  been appropriately renamed, e.g.:
\bea
c0(m_1^2,m_2^2,m_1^2) \to c0new(m_1^2,m_2^2)
\label{eq:repl}
\eea
\end{itemize}

To give an example, lets consider the 3-point scalar-fermion-fermion
1-loop amplitude, the triangle diagram with Dirac fermions $C_n$,
Majorana fermions $N_j$ and scalars $D_i$ circulating in the loop.  In
general such an amplitude is the sum of terms depending on mixing
matrices and 3-point loop function, one of them could look like
\bea
Amplitude = Z_D^{Ii} Z_D^{Ji*} O_N^{Kj} O_N^{Lj*} V_C^{Mn*} U_C^{Nn}
m_{C_n} c_{21}(p,q,m_{C_n}^2,m_{D_i}^2,m_{N_j}^2)
\eea
To expand this in terms of mass insertions with the use of \MMI{}
routines, one must rewrite it into a Mathematica expression as:\\[2mm]
{\small \tt
\begin{tabular}{ll}
Amplitude = & SMIX[D,I,i] Conjugate[SMIX[D,J,i]] \\
& NMIX[N,K,j] Conjugate[NMIX[N,L,j]] \\
& Conjugate[FMIXL[C,M,n]] FMIXR[C,N,n] MASS[C,n] \\
& LOOP[c21,\{\{C,n\},\{D,i\},\{N,j\}\},\{p,q\}];\\[3mm]
\end{tabular}
}

\subsection{Control variables in \MMI}

\MMI{} package uses the following control variables, defining which
combinations of mixing matrices should be expanded to which order in
terms of mass insertions, and in which form the final result should be
presented.

\begin{itemize}
\item {\tt \bf FetScalarList}={\tt \{\{P1,o1\},\ldots,\{Pn,on\}\}}.
  In this list the user can specify symbols for the scalar and vector
  particles ({\tt P1,\ldots,Pn}) for which the mixing matrices are
  expanded in MI powers to orders {\tt o1,\ldots,on}, respectively.
\item {\tt \bf FetFermionList}={\tt
  \{\{P1,o1,M1\},\ldots,\{Pn,on,Mn\}\}}.  Specifies list of Dirac or
  Majorana fermions {\tt P1,\ldots,Pn} for which the mixing matrices
  are expanded to MI orders {\tt o1,\ldots,on}.  Each of the arguments
  {\tt M1,\ldots,Mn} can take one of two values, {\tt MHM} or {\tt
    MMH}, deciding which form of the RHS in the last 2 lines
  of~\eqs{eq:fermionFET}{eq:majoranaFET} is used, depending on $M^\dag
  M$ or on $MM^\dag$, respectively (see discussion in
  Section~\ref{sec:ambi}).
\item {\tt \bf FetMaxOrder=n}.  Only mass insertion products of the
  total order {\tt FetMaxOrder} or lower are kept in the final result.
%
%
\end{itemize}
Particles with symbols not specified on {\bf \tt FetScalarList} or
{\bf \tt FetFermionList} lists are left unexpanded.

\subsection{The main expansion routine and output syntax}
\label{sec:out}

The actual MI expansion is done by a call to the function \\[4mm]
{\tt \bf FetExpand[ Amplitude ] } \\[4mm]
The {\tt FetExpand} routine performs the following actions:
\begin{enumerate}
\item It checks first the structure of the amplitude for possible
  syntax problems, displaying if necessary the relevant error or
  warning messages.
\item It finds pairs of mixing matrices for which the {\em second}
  index is identical (like e.g. in $X_Q^{Ai*} X_Q^{Bi}$ product), and
  the particle symbol ($Q$ in this case) appears on {\tt
    FetScalarList} or on {\tt FetFermionList} (as mentioned above it
  is assumed that the repeating index denotes summation over it).  For
  pairs of fermionic matrices, the program also searches if they are
  associated with any linear fermion mass factors with the same
  summation index.
\item The routine finds the loop functions which arguments have
  indices identical to those in the matching pairs of mixing matrices
  and performs the FET expansion for each of such pairs, using the
  formula of~\eqst{eq:scalarFET}{eq:majoranaFET}
  and~\eq{theorem}. Some further syntax checks are performed at this
  stage and reported if necessary.
\item Finally the result is displayed in terms of the objects used to
  denote diagonal and off-diagonal entries of the mass matrices listed
  in Table~\ref{tab:final}.
\end{enumerate}

\begin{table}[tb]
\begin{tabular}{lll}
\hline
Object & Syntax and arguments & Example\\
\hline
Diagonal entry of scalar & {\tt MS2[P,i,i]} & $(M_P^2)_{ii} =$ {\tt
  MS2[P,i,i]} \\
squared mass matrix & $P$ - particle symbol & \\
& $i$ - particle index\\[3mm]
Fermion mass matrix & {\tt MF[P,i,j]} &
  $(M_P)_{ij} =$ {\tt MF[P,i,j]} \\
& $P$ - particle symbol & \\
& $i,j$ - particle indices\\[3mm]
Diagonal entry of squared $M^\dagger M$ & {\tt MHM[P,i,i]} &
  $(M_P^\dagger M_P)_{ii} =$ {\tt MHM[P,i,i]} \\
fermion mass matrix & $P$ - particle symbol & \\
& $i$ - particle index\\[3mm]
Diagonal entry of squared $MM^\dagger$ & {\tt MMH[P,i,i]} & $(M_P
  M_P^\dagger)_{ii} =$ {\tt MMH[P,i,i]} \\
fermion mass matrix & $P$ - particle symbol & \\
& $i$ - particle index\\[3mm]
Scalar MI matrix & {\tt MS2I[P,i,j]} & $(M_P^2)_{ij} =$ {\tt
  MS2I[P,i,j]} \\
  & $P$ - particle symbol & \\
& $i,j$ - particle indices\\[3mm]
Fermion squared $M^\dagger M$ MI matrix & {\tt MHMI[P,i,j]} &
  $(M_P^\dagger M_P)_{ij} =$ {\tt MHMI[P,i,j]} \\
 & $P$ - particle symbol & \\
& $i,j$ - particle index\\[3mm]
Fermion squared $MM^\dagger$ MI matrix  & {\tt MMHI[P,i,i]} & $(M_P
  M_P^\dagger)_{ij} =$ {\tt MMHI[P,i,j]} \\
 & $P$ - particle symbol & \\
& $i,j$ - particle index\\[3mm]
\hline
\end{tabular}
\caption{Notation for the mass matrices in the interaction basis
  appearing in the expanded amplitudes. The objects denoting
  off-diagonal mass insertions, {\tt MS2I, MHMI} and {\tt MMHI}, are
  assumed to have 0's on diagonal. \label{tab:final}}
\end{table}

The expanded amplitude depends on the divided differences of the loop
functions appearing in the original unexpanded expression. The
notation for the divided differences is as follows:
\begin{itemize}
\item The loop functions are renamed using the first argument of the
  original expression; the new first argument (non-negative integer)
  is added to denote the order of divided difference. The list
  containing the other arguments is flattened, so that the arguments
  are no longer splitted into masses and optional variables. This is
  illustrated by the following example:\\[2mm]
{\small \tt LOOP[b21,\{\{U,i\},\{C,j\}\},\{p\}] $\to$
  b21[0,\{U,i\},\{C,j\},p ] }
\item The physical mass arguments of the loop functions undergoing the
  MI expansion are replaced by a list containing the diagonal entries
  of mass matrices (boson or fermion squared masses) representing the
  variables of divided difference in the given argument.
\item The divided differences of the loop functions are multiplied by
  the relevant mass insertion factors. Again, it is assumed that all
  repeated indices named {\tt fetQxx} should be summed over.

Continuing the example given above, lets assume that MI expansion has
been performed to 2nd order for a fermion $C$. Then the highest order
term in the expansion is proportional to the function denoted
as:\\[2mm]
{\small \tt Conjugate[FMIXL[C,c,j]] FMIXL[C,d,j]
  LOOP[b21,\{\{U,i\},\{C,j\}\},\{p\}]} $\to$ \\[2mm]
{\small \tt MMHI[C,d,fetC1] MMHI[C,fetC1,c] $\times$ \\
$\times$ b21[2,\{U,i\},\{MMH[C,d,d],MMH[C,fetC1,fetC1],MMH[C,c,c]\},p]
} \\[2mm]
where $2$ being used as the 1st argument of the {\tt b21} function is
the order of the divided difference, {\tt MMH[C,i,i]} = $(M_C
M_C^\dag)_{ii}$, {\tt MMHI[C,i,j]} = $(M_C M_C^\dag)_{ij}$ for $i\ne
j$ and the summation over the internal index {\tt fetC1} is assumed.
\item If the FET expansion is performed in several indices, the first
  argument of the expanded loop function denotes the total order (the
  sum of all orders) of the divided differences in all arguments.
\item If the final result still depends on some of the physical masses
  (not expanded into MI series), they are now denoted as {\tt
  MASS[P,i]}, e.g.\\[2mm]
{\small \tt
  b21[2,\{U,i\},\{MMH[C,d,d],MMH[C,fetC1,fetC1],MMH[C,c,c]\},p] $\to$
  \\[2mm]
\hspace*{1cm}
b21[2,MASS[U,i]$^2$,\{MMH[C,d,d],MMH[C,fetC1,fetC1],MMH[C,c,c]\},p] }
\item If necessary, the divided differences can be re-expressed as the
  explicit combinations of the original loop functions with the use of
  {\tt FetExpandDividedDifferences} functions described in next
  section.
\end{itemize}

\subsection{Auxiliary functions}
\label{sec:aux}

In addition to the main {\tt FetExpand} routine, the \MMI{} package
provides several auxiliary functions. For consistency and to
distinguish them from other routines provided by Mathematica, all
their names, by convention, start from the prefix {\tt Fet}.

One of the auxiliary functions, {\tt FetExpandDividedDifferences},
allows to re-express divided differences appearing in the {\tt
  FetExpand} output as the combinations of the initial loop
functions. Other functions help to manipulate expressions with
repeating indices, assumed to be implicitly summed over.  They
naturally appear in the higher order terms of the FET expansion
(see~\eq{theorem}) in sums over mass matrix indices.  In many cases,
the summation convention can be also used to define the initial
(unexpanded) amplitude in a compact form.  However, for the analysis
of the physical effects it may be useful to expand some (or all)
repeated indices as explicit sums.  For that purpose, \MMI{} package
provides 3 routines, {\tt FetSumParticle}, {\tt FetSumFactor} and {\tt
  FetSumIndex}.  In addition, Mathematica, at least in version 10,
does not have summation convention rules with Kronecker
$\delta$-symbol implemented.  Resumming Kronecker symbols can be done
using the {\tt FetSumWithDelta} function.

Below we describe in details the syntax for using the auxiliary
functions (their compact list is collected in Table~\ref{tab:aux}).
One should note that some of their arguments, given in parenthesis
with slanted font, are optional.

\begin{table}[tb]
{\small
\begin{tabular}{p{15mm}ll}
\hline
Function & Arguments \\
\hline\\
\multicolumn{3}{l}{\tt FetExpandDividedDifferences[ DivFunction, {(\em
      ReplacementRule)} ]}\\
& {\tt DivFunction} & divided difference\\
& {\tt ReplacementRule} & rule how to redefine Function (optional)\\[3mm]
\multicolumn{3}{l}{\tt FetSumWithDelta[ Expression, {(\em
      ExcludeList)} ]}\\
& {\tt Expression} & expression containing Kronecker $\delta$-symbols\\
& {\tt ExcludeList} & index or list of indices excluded from summation
(optional)\\[3mm]
\multicolumn{3}{l}{\tt FetSumParticle[ Expression, Particle, Range,
    {(\em ExcludeList)} ]}\\
& {\tt Expression} & expression to expand repeating indices \\
& {\tt Particle} & symbol of particle (or list of particles) which
indices are expanded\\
& {\tt Range=\{Low,Up\} } & summation range for repeating indices \\
& {\tt ExcludeList} & index or list of indices excluded from summation
(optional)\\[3mm]
\multicolumn{3}{l}{\tt FetSumFactor[ Expression, Factor, Range, {(\em
      ExcludeList)} ]}\\
& {\tt Expression} & expression to expand repeating indices \\
& {\tt Factor} & factor name (or list of factors) which indices are
expanded\\
& {\tt Range=\{Low,Up\} } & summation range for repeating indices \\
& {\tt ExcludeList} & index or list of indices excluded from summation
(optional)\\[3mm]
\multicolumn{3}{l}{\tt FetSumIndex[ Expression, Index, Range ]}\\
& {\tt Expression} & expression to expand  \\
& {\tt Index} & name of index to expand \\
& {\tt Range=\{Low,Up\} } & summation range for {\tt Index} \\[3mm]
\hline
\end{tabular}
\caption{List and arguments of auxiliary functions provided by \MMI{}
  package. \label{tab:aux}} }
\end{table}

\subsubsection{\tt \bf FetExpandDividedDifferences[ DivFunction, {(\em
      ReplacementRule)} ]}

{\tt FetExpandDividedDifferences} expands the divided difference of a
function given~as\\[2mm]
{\small \tt DivFunction = fname[(n +
    \ldots,\{M\_1,\ldots,M\_(n+1)\},$\ldots$,optional~arguments]
}\\[2mm]
into an explicit combination of the initial functions with different
mass arguments (the first argument, order of divided difference, is
truncated in the expanded form).  For the degenerate mass arguments
relevant derivatives (in the standard Mathematica notation) of the
initial function are used. Examples are given below.\\[1mm]
Non-degenerated arguments:\\[2mm]
{\small \tt
\begin{tabular}{l}
FetExpandDividedDifferences[ b21[2,\{m1,m2,m3\},M1,p] ] =\\[2mm]
~~~ b21[m1,M1,p]/(m1 - m2)/(m1 - m3) - b21[m2,M1,p]/(m1 - m2)/(m2 - m3) \\[2mm]
~ + b21[m3,M1,p]/(m1 - m3)/(m2 - m3)
\end{tabular}\\[2mm]
}
Degenerated arguments:\\[2mm]
{\small \tt
\begin{tabular}{l}
FetExpandDividedDifferences[ b21[2,\{m1,m2,m1\},M1,p] ] =\\[2mm]
~~~ (b21[m2,M1,p] - b21[m1,M1,p])/(m1 - m2)$^2$ +
b21$^{(1,0,0)}$[m2,M1,p]/(m1 - m2)\\[2mm]
\end{tabular}\\[2mm]
\begin{tabular}{l}
FetExpandDividedDifferences[ b21[2,\{m1,m1,m1\},M1,p] ] =
b21$^{(2,0,0)}$[m1,M1,p]/2\\[2mm]
\end{tabular}\\[2mm]
}
Divided difference of multiple arguments, of total order
$2+1=3$:\\[2mm]
{\small \tt
\begin{tabular}{l}
FetExpandDividedDifferences[ b21[3,\{m1,m1,m1\},\{M1,M2\},p] ] =\\[2mm]
~~~ (b21$^{(2,0,0)}$[m1,M1,p] - b21$^{(2,0,0)}$[m1,M2,p])/2/(M1 -
M2)\\[2mm]
\end{tabular}\\[2mm]
}

The second optional argument of {\tt FetExpandDividedDifferences}
should be specified if the function used as the first argument was
renamed to avoid multiple identical mass arguments, as in the example
given in~\eq{eq:repl}.  Specifying the replacement rule as the second
argument allows to revert the redefinition and to go back to the
original function name, argument list and, for the degenerate mass
arguments, to calculate correctly the function derivatives.  Example:
\\[3mm]
{\small \tt
\begin{tabular}{l}
FetExpandDividedDifferences[ c0new[2,\{m1,m2,m1\},M1] ] =\\[2mm]
~~~~~~~~~~~~~~ (c0new[m2,M1] - c0new[m1,M1])/(m1 - m2)$^2$ \\[2mm]
~~~~~~~~~~~~ + c0new$^{(1,0)}$[m1,M1]/(m1 - m2)
\end{tabular}\\[3mm]
}
but\\[3mm]
{\small \tt
\begin{tabular}{l}
FetExpandDividedDifferences[ c0new[2,\{m1,m2,m1\},M1],
  c0new[a\_,b\_]->c0[a,b,a] ] =\\[2mm]
~~~~~~~~~~~~~~ (c0[m2,M1,m2] - c0[m1,M1,m1])/(m1 - m2)$^2$ \\[2mm]
~~~~~~~~~~~~ + (c0$^{(1,0,0)}$[m1,M1,m1] + c0$^{(0,0,1)}$[m1,M1,m1])/(m1 - m2)
\end{tabular}
}

\subsubsection{\tt \bf FetSumWithDelta[ Expression, {(\em ExcludeList)} ]}

{\tt FetSumWithDelta} function searches for occurrences of the factor
{\tt KroneckerDelta[a,b]} in {\tt Expression} and if possible replaces
them using the standard rule\\[2mm]
{\tt KroneckerDelta[a,b] X[a] -> X[b]},\\[2mm]
excluding the cases when summation index is specified on the
(optional) {\tt ExcludeList}. This option allows e.g. to exclude
external particle indices from summation.  Example:\\[2mm]
{\small \tt exp = KroneckerDelta[I,a] KroneckerDelta[J,b] Fun[a,b]
  \\[2mm]
\hspace*{1cm} FetSumWithDelta[ exp ] = Fun[I,J]\\[2mm]
\hspace*{1cm} FetSumWithDelta[ exp, \{b\} ] = KroneckerDelta[J,b]
Fun[I,b]\\[2mm]
}
Few additional remarks are in order:
\begin{itemize}
\item {\tt FetSumWithDelta} does not perform summation if the
  repeating index is not a simple variable but an expression by
  itself, so terms of the form\\[2mm]
{\small \tt KroneckerDelta[a+1,b] f[a]}\\[2mm]
are left unchanged. However the argument of {\tt f} is evaluated
properly:\\[2mm]
{\small \tt FetSumWithDelta[ KroneckerDelta[a,b] f[a+1] ] = f[b+1]}.
\item If the second {\tt KroneckerDelta} argument, complementary to
  the summation index, is an expression, replacement is done:\\[2mm]
{\tt FetSumWithDelta[ KroneckerDelta[a,b+3] f[a] ] = f[b+3]}\\[2mm]
However, {\tt FetSumWithDelta} does not check for the allowed range
for summation indices - if in the example above {\tt a,b = 1...3},
then {\tt KroneckerDelta[a,b+3] $\equiv$ 0} and the result given by
{\tt FetSumWithDelta} is not correct. Expressions of that type must be
simplified manually by the user.
\end{itemize}

\subsubsection{\tt \bf FetSumParticle[ Expression, Particle, Range, {\em
      (ExcludeList)} ]}

{\tt FetSumParticle} expands the summation convention for the
repeating indices of mass insertion matrices for given set of
particles specified in the argument {\tt Particle} (it could be a
single particle symbol or list of symbols). More specifically, it
searches for the occurrence of factors of the form {\tt
  X[Particle,a,b]}, where {\tt X} is one of the following matrices:
{\tt X=MF,MS2I,MHMI,MMHI}.  If such a factor is found, and at least
one of its indices (not specified on the optional argument {\tt
  ExcludeList}) is present also in the term multiplying it, the
repeating index is replaced by the explicit sum in the range {\tt
  Range=\{Low,Up\}}.  Examples:\\[2mm]
{\small \tt exp = MF[Q,B,K] MF[Q,B,J] + MS2I[P,K,A] MA[A,J] 
 \\[2mm]
FetSumParticle[ exp,P,\{1,2\} ] =
MF[Q,B,J] MF[Q,B,K] \\[2mm]
\hspace*{20mm} + MA[1,J] MS2I[P,K,1] + MA[2,J] MS2I[P,K,2] \\[2mm]
FetSumParticle[ exp,\{P,Q\},\{1,2\} ] = MF[Q,1,J] MF[Q,1,K]\\
\hspace*{20mm} + MF[Q,2,J] MF[Q,2,K] + MA[1,J] MS2I[P,K,1] + MA[2,J]
MS2I[P,K,2]\\[2mm]
} 
Note that whenever possible, {\tt FetSumParticle} (and also other
functions dealing with summation convention described below)
automatically sets to zero diagonal elements of the mass insertion
matrices, like {\tt MS2I[P,2,2]} factor in the example below:\\[2mm]
{\small \tt FetSumParticle[ exp /.~K->2 ,\{P,Q\},\{1,2\} ] = MF[Q,1,J]
  MF[Q,1,2]\\
\hspace*{20mm} + MF[Q,2,J] MF[Q,2,2] + MA[1,J] MS2I[P,2,1]
} 

\subsubsection{\tt \bf FetSumFactor[ Expression, Factor, Range, {\em
      (ExcludeList)} ]}

{\tt FetSumFactor} routine can be used to expand summation over
indices of objects not related to MI expansion (like CKM matrix).  It
searches for the repeating indices of factor(s) specified as the
argument {\tt Factor}. Factors can have any number of indices, but
they are all summed in the same {\tt Range=\{Low,Up\}}.  If various
indices have different summation ranges, routine should be called
several times, specifying optional {\tt ExcludeList} argument to avoid
summation over indices not belonging to correct {\tt Range}.
Examples:\\[2mm]
{\small \tt exp = MA[A,K] CKM[A,J] \\[2mm]
FetSumFactor[ exp,CKM,\{1,3\} ] = CKM[1,J] MA[1,K] + CKM[2,J]
MA[2,K]\\
+ CKM[3,J] MA[3,K] \\[2mm]
FetSumFactor[ exp,CKM,\{1,3\}, A ] = CKM[A,J] MA[A,K]
}

\subsubsection{\tt \bf FetSumIndex[ Expression,  Index, Range ] }

{\tt FetSumIndex} routine can be used to expand summation over single
index directly specified by user.  Example:\\[2mm]
{\small \tt exp = 1 + A[J,K] B[J,L] \\[2mm]
FetSumIndex[ exp,J,\{1,3\} ] = 1 + A[1,K] B[1,L] + A[2,K] B[2,L] +
A[3,K] B[3,L]
}

\subsubsection{Special cases and limitations}

Syntax of \MMI{} package assumes in general that repeating indices are
implicitly summed over.  However, this is not a summation convention
in a ``classical'' sense, as it allows for indices repeating more than
twice.  Such definition is convenient for applications of FET
expansion, but, used without proper care, may lead to various
ambiguities. Thus, it is advisable to check the structure of the
intermediate expressions appearing during calculations, particularly
before applying to them auxiliary functions designed for manipulating
objects with repeating indices.

Few additional remarks may be helpful to avoid problems with the
incorrect usage of routines provided by \MMI{} package:
\begin{itemize}
\item It is highly recommended to use the {\tt ExcludeList} argument
  of {\tt FetSumWithDelta, FetSumParticle} and {\tt FetSumFactor}
  functions for specifying which indices correspond to external
  particles in given amplitude, to avoid resumming over them
  accidentally.
\item Notice that the functions {\tt FetSumFactor, FetSumParticle} and
  {\tt FetSumWithDelta} search only for repeating indices in simple
  factors in the analyzed expression, not checking for their
  appearance in lower level sub-expression (those are left
  unchanged). For example:\\[3mm]
{\small \tt FetSumWithDelta[ KroneckerDelta[a,b] f[a] ] = f[b]}\\[2mm]
but\\[3mm]
{\small \tt FetSumWithDelta[ Fun[KroneckerDelta[a,b] f[a]] ] =\\
\hspace*{5cm}  Fun[f[a] KroneckerDelta[a,b]] }\\[3mm]
and similarly for {\tt FetSumParticle} and {\tt FetSumFactor}.  {\tt
  FetSumIndex} will work even for indices hidden in sub-expressions,
assuming that they appear in at least two different factors.
\item Functions {\tt FetSumParticle} and {\tt FetSumFactor} treat
  higher powers of matrix elements as products with repeating indices,
  and perform appropriate resummations, as illustrated below:\\[3mm]
{\small \tt FetSumFactor[ M[a,b]{\textasciicircum}2,M,\{1,2\} ]
  $\equiv$ FetSumFactor[ M[a,b]M[a,b],M,\{1,2\} ] = \\
\hspace*{5cm} M[1,1]$^2$ + M[1,2]$^2$ + M[2,1]$^2$ + M[2,2]$^2$}
\item One should strictly avoid using the same names for various types
  of quantities appearing in the transition amplitudes or other
  expressions used as arguments of \MMI{} functions. In particular,
  functions dealing with summation convention cannot distinguish what
  is really an ``index'' and try to sum over all symbols in given
  expression with the names identical to repeating indices. This may
  lead to strange bugs and obviously incorrect or nonsensical results,
  like in the examples given below:\\[2mm]
{\tt FetSumWithDelta[ KroneckerDelta[A,B] A[A] ] = B[B]}\\[2mm]
{\tt FetSumFactor[ M[A,B] A[A],A,\{1,2\} ] = 1[1] M[1,B] + 2[2] M[2,B]
}
\end{itemize}

\section{Examples of the \MMI{} applications}
\setcounter{equation}{0}
\label{sec:example}

We collect below several examples illustrating the functionality of
the \MMI{} package, from simple test amplitudes to the realistic case
of the expansion of one of the diagrams contributing to the flavor
violating decay of the Higgs boson to leptons, $h\to \tau\mu$, in the
MSSM.

\subsection{Expansion of simple amplitudes}

The two basic examples of \MMI{} usage presented below are included in
the file {\tt mmi\_example.m} attached to the \MMI{} distribution.\\

\noindent {\bf Example 1.}  The simplest case involves a pair of mixing
matrices of a single scalar particle multiplying loop function
$a(m^2)$.  Lets assume that amplitude should be expanded to 2nd order
in MI powers. \\[2mm]
{\small \tt
Ampl = Conjugate[SMIX[P,a,i]] SMIX[P,b,i] LOOP[A0,\{\{P,i\}\}];\\[1mm]
FetScalarList = \{\{P,2\}\};\\[1mm]
FetMaxOrder = 2;\\[1mm]
FetAmpl = FetExpand[Ampl];\\[1mm]
Print["Expanded amplitude = ", FetAmpl];\\[3mm]
}
The result is:\\[2mm]
{\small \tt
Expanded amplitude = A0[0,MS2[P,b,b]] KroneckerDelta[a,b] + \\[1mm]
A0[1,\{MS2[P,b,b],MS2[P,a,a]\}] MS2I[P,b,a] + \\[1mm]
A0[2,\{MS2[P,b,b],MS2[P,fetP1,fetP1],MS2[P,a,a]\}] MS2I[P,b,fetP1]
MS2I[P,fetP1,a]\\[3mm]
}
Note that only amplitudes of a specific structure
(see~\eqst{eq:scalarFET}{eq:majoranaFET}) are allowed and {\tt
  FetExpand} reports errors if incorrect combinations of mixing
matrices are used.  For example, if c.c. is removed from one of the
scalar matrices in the example above, \\[2mm]
{\small \tt
Ampl = SMIX[P,a,i] SMIX[P,b,i] LOOP[A0,\{\{P,i\}\}];\\[3mm]
}
program reports:\\[2mm]
{\small \tt
MassToMI::cmplx2: ERROR: Incorrect matching of c.c.'s in product of
SMIX[P,a,i] and SMIX[P,b,i]?\\[3mm]
}
and aborts the execution.

In other cases only warnings or informational messages are displayed
but program still makes an attempt to calculate the result, like for
instance the mismatch in the summation index of scalar matrices:
\\[2mm]
{\small \tt
Ampl = Conjugate[SMIX[P,a,j]] SMIX[P,b,i] LOOP[A0,\{\{P,i\}\}];\\[3mm]
}
leads to\\[2mm]
{\small \tt
MassToMI::war1: WARNING: unmatched index of scalar on FetExpandList in\\
SMIX[P,a,j]?\\[1mm]
MassToMI::war1: WARNING: unmatched index of scalar on FetExpandList in\\
SMIX[P,b,i]?\\[3mm]
Expanded amplitude = A0[0,MASS[P,i]$^2$] Conjugate[SMIX[P,a,j]]
SMIX[P,b,i] \\[3mm]
}
meaning that no pair of scalar mixing matrices with matching 2nd index
was found, so the result is simply equal to the initial amplitude
rewritten using the syntax rules for the \MMI{} output.

The full list of errors, warnings and informational messages can be found
in the header of the {\tt MassToMI.m} file.\\[3mm]

\noindent {\bf Example 2.} Let us now consider a more complicated case
of an amplitude depending on scalar $P$ and Dirac fermion $F$
masses. We request 1st order of MI expansion for both particles and
set {\tt FetMaxOrder = 1} as a maximum total MI order. The preferred
form of the squared fermion mass in output is $M^\dag M$.\\[2mm]
{\small \tt
Ampl = Conjugate[SMIX[P,a,i]] SMIX[P,b,i] Conjugate[FMIXL[F,c,j]]
FMIXR[F,d,j] MASS[F,j] LOOP[B0,\{\{P,i\},\{F,j\}\}];\\[1mm]
FetScalarList = \{\{P,1\}\};\\[1mm]
FetFermionList = \{\{F,1,MHM\}\};\\[1mm]
FetMaxOrder = 1;\\[1mm]
FetAmpl = FetExpand[Ampl];\\[1mm]
Print["Expanded amplitude = ", FetAmpl];\\[3mm]
}
The result is\\[2mm]
{\small \tt
Expanded amplitude = B0[0,MS2[P,b,b], MHM[F,d,d]]
Conjugate[MF[F,c,fetF1]]\\
KroneckerDelta[a,b] KroneckerDelta[d,fetF1] \\
+ B0[1,MS2[P,b,b],\{MHM[F,d,d],MHM[F,fetF1,fetF1]\}]
Conjugate[MF[F,c,fetF1]] \\
KroneckerDelta[a,b] MHMI[F,d,fetF1] \\
+ B0[1,\{MS2[P,b,b],MS2[P,a,a]\},MHM[F,d,d]] Conjugate[MF[F,c,fetF1]] \\
KroneckerDelta[d,fetF1] MS2I[P,b,a] \\[3mm]
}
where again summation over the repeating index {\tt fetF1} is
assumed. The result contains some Kronecker $\delta$-symbols with
fermion indices which can be resummed\\[3mm]
{\small \tt
FetAmpl = FetSumWithDelta[FetAmpl,\{a,b,c,d\}];\\[2mm]
}
giving the simpler expression\\[2mm]
{\small \tt
Expanded amplitude = B0[0,MS2[P,b,b],MHM[F,d,d]] Conjugate[MF[F,c,d]] \\
KroneckerDelta[a,b] \\
+ B0[1,MS2[P,b,b],\{MHM[F,d,d],MHM[F,fetF1,fetF1]\}]
Conjugate[MF[F,c,fetF1]]\\
KroneckerDelta[a,b] MHMI[F,d,fetF1] \\
+ B0[1,\{MS2[P,b,b],MS2[P,a,a]\},MHM[F,d,d]] Conjugate[MF[F,c,d]]
MS2I[P,b,a]
}

\subsection{$h\to \tau\mu$ decay in the MSSM.}

As a more realistic and complicated example we consider the flavor
violating decay of the Higgs boson to leptons, $h\to \tau\mu$, in the
MSSM.  This decay was recently measured by ATLAS~\cite{Aad:2015gha}
and CMS~\cite{Khachatryan:2015kon} Collaborations, and reported to be
more frequent than predicted within the Standard Model. It was also
theoretically recently reanalyzed within the
MSSM~\cite{Arana-Catania:2013xma} and claimed to be enhanced.

\begin{figure}[htbp]
\begin{center}
\begin{picture}(130,120)(10,-15)
\ArrowLine(10,0)(35,0)
\Text(-5,0)[l]{$l^B$}
\ArrowLine(105,0)(130,0)
\Text(145,0)[c]{$l^A$}
\DashLine(70,60)(70,100){4}
\Text(75,100)[l]{$H_0^K$}
\ArrowLine(35,0)(105,0)
\DashLine(35,0)(70,60){4}
\DashLine(105,0)(70,60){4}
\Text(110,25)[c]{$L_l$}
\Text(35,25)[c]{$L_i$}
\Text(70,-10)[c]{$\chi_0^j$}
\Vertex(35,0){2}
\Vertex(105,0){2}
\Vertex(70,60){2}
\end{picture}
\caption{Slepton-neutralino diagram contributing to the $h\to\tau\mu$
decay in the MSSM\label{fig:diag}}
\end{center}
\end{figure}
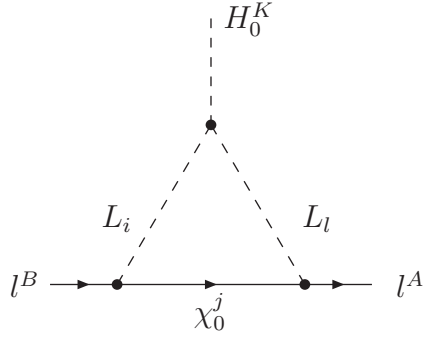

For simplicity, for the purpose of our example we expand the
contribution from one diagram only, shown in Fig.~\ref{fig:diag}
(results for MI expansion of sum of all diagrams contributing to the
effective Higgs-fermion vertex in the MSSM can be found in
Refs.~\cite{Crivellin:2010er, Crivellin:2011jt, Crivellin:2012zz}). It
depends on the scalar (slepton) and fermionic (neutralino) mixing
matrices.

In the approximation of the vanishing external momenta, the amplitude
corresponding to this diagram reads as:
\bea
{\cal A} = \Gamma_L^{ABK} P_L + \Gamma_R^{ABK} P_R \;.
\label{eq:gamdef}
\eea
where
\bea
\Gamma_{L(R)}^{ABK} &=& -\sum_{i,j,l}
\,m_{\chi_{j}}(V_{HLL}^{Kli}V_{lL\chi,R(L)}^{Alj\,*}
V_{lL\chi,L(R)}^{Bij}) \, \,C_{0}
[m_{L_i},m_{L_l},m_{\chi_j}]\;, \label{CSFS}
\eea
and $A,B$ are lepton generation indices ($B=3$ and $A=2$ for $\tau\mu$
in the final state, $K$ is the Higgs index, $K=2$ for the ``little
$h$'' Higgs boson and $K=1$ for the ``big $H$'' and $C_0$ is the
standard 3-point scalar loop integral for vanishing external momenta.

Following strictly the notation and conventions of
Refs.~\cite{Rosiek:1989rs, Rosiek:1995kg} (we do not repeat them here
explicitly), the Higgs-slepton and lepton-slepton-neutralino vertices
read as
\bea
V_{lL\chi,L}^{Aij} &=& {e \over \sqrt{2}s_Wc_W} Z_L^{Ai} (Z_N^{1j} s_W
+ Z_N^{2j} c_W) + Y_l^A Z_L^{(A+3)i} Z_N^{3j} \\
V_{lL\chi,R}^{Aij} &=& {-e\sqrt{2} \over c_W} Z_L^{(A+3)i}
Z_N^{1j\star} + Y_l^A Z_{L}^{Ai} Z_{N}^{3j\star}\\
V_{HLL}^{Kli} &=& \sum_{C=1}^3 \left( {e^2 \over 2c_W^2} \left( v_1
Z_R^{1K} - v_2 Z_R^{2K} \right) \left(\delta^{il} + {1-4s_W^2 \over
2s_W^2} Z_L^{Ci\star} Z_L^{Cl}\right) \right.  \nonumber\\
& - &  (Y_l^C)^2 v_1 Z_R^{1K} (Z_L^{Ci\star} Z_L^{Cl} +
 Z_L^{(C+3)i\star} Z_L^{(C+3)l}) \nonumber\\
&-& \left. {Z_R^{2K}\over \sqrt{2}} Y_l^{C}  (\mu^{\star} Z_L^{Ci\star}
Z_L^{(C+3)l} + \mu Z_L^{Cl} Z_L^{(C+3)i\star})\right)\nonumber\\
&-& {1\over \sqrt{2}} \sum_{C,D=1}^3 \left( Z_R^{1K} (A_l^{CD\star} Z_L^{Cl}
Z_L^{(D+3)i\star} + A_l^{CD} Z_L^{Ci\star} Z_L^{(D+3)l} ) \right. \nonumber\\
& -& \left. Z_R^{2K} (A_l^{'CD\star} Z_L^{Cl} Z_L^{(D+3)i\star} +
A_l^{'CD} Z_L^{Ci\star} Z_L^{(D+3)l}) \right)
\label{eq:hll}
\eea
Neglecting the small terms proportional to lepton Yukawa couplings
and, for simplicity, the ``non-holomorphic'' $A_l'$ trilinear soft
terms, the vertices and amplitude for $\Gamma_L^{ABK}$ are coded as
(summation over repeating generation indices {\tt C,D} in {\tt HLL} is
assumed)\footnote{The Mathematica code for the example discussed in
  this section is attached to the \MMI{} distribution as the file {\tt
    htaumu\_decay.m}}:\\[2mm]
{\small \tt
HLL[K\_,l\_,i\_] = - e{\textasciicircum}2/2/cw{\textasciicircum}2 (v1
ZR[1,K] - v2 ZR[2,K]) (KroneckerDelta[i,l] \\
+ (1 - 4 sw{\textasciicircum}2)/2/sw{\textasciicircum}2
Conjugate[SMIX[L,C,i]] SMIX[L,C,l])\\
- 1/Sqrt[2] ZR[1,K] (Conjugate[AL[C,D]] SMIX[L,C,l]
Conjugate[SMIX[L,D+3,i]] + AL[C,D] Conjugate[SMIX[L,C,i]]
SMIX[L,D+3,l]);\\[3mm]
LSNL[J\_,j\_,i\_] = e/(Sqrt[2] sw cw) SMIX[L,J,j] (NMIX[N,1,i] sw +
NMIX[N,2,i] cw);\\
LSNR[J\_,j\_,i\_] =- e Sqrt[2]/cw SMIX[L,J+3,j]
Conjugate[NMIX[N,1,i]];\\[3mm]
(* amplitude *)\\
ampl = - MASS[N,j] HLL[K,l,i] Conjugate[LSNR[A,l,j]] LSNL[B,i,j] *\\
~~~~~~~~~LOOP[c0,\{\{L,i\},\{L,l\},\{N,j\} \}] // Expand;\\[3mm]
(* simplification: sum over KroneckerDelta[i,l] *)\\
ampl = FetSumWithDelta[ampl,\{A,B,K\}];\\
(* give new name for loop function with repeating arguments *)\\
ampl = ampl /. LOOP[c0,\{\{L,i\_\},\{L,i\_\},\{N,j\}\}] ->
LOOP[c0sq,\{\{L,i\},\{N,j\}\}];\\[2mm] }
At this stage the amplitude is correctly defined in terms of \MMI{}
objects (expressions for $\Gamma_R^{ABK}$ can be obtained by replacing
$L\leftrightarrow R$).  For our purpose we expand it up to the lowest
non-trivial order, i.e. 1st order in the slepton mass insertion and
0th order in the neutralino mass insertion (neutralino mass matrix
entries are not related to flavor violation, so neglecting higher
orders is equivalent to skipping ${\cal O}(M_W^2/M_{SUSY}^2)$
suppressed terms). The actual expansion is done by:\\[3mm]
{\small \tt (* control variables *)\\
FetMaxOrder = 1;\\
FetScalarList = \{\{L,1\}\};\\
FetFermionList = \{\{N,0,MHM\}\};\\[2mm]
(* MAIN EXPANSION ROUTINE *)\\
fetampl = FetExpand[ ampl ];\\[2mm]
}
The expanded amplitude is a lengthy expression build from objects
defined in Section~\ref{sec:out}.  It could be further significantly
simplified by substituting the explicit form for the neutralino and
slepton mass matrices. \\[3mm]
{\small \tt
(* simplify KroneckerDelta terms, excluding external indices *)\\
fetampl = fetampl /. KroneckerDelta[3 + a\_, 3 + b\_] -> KroneckerDelta[a, b];\\
fetampl = fetampl /. KroneckerDelta[3 + a\_, b\_] -> 0;\\
fetampl = fetampl /. KroneckerDelta[a\_, 3 + b\_] -> 0;\\
fetampl = FetSumWithDelta[ fetampl, \{A,B,K\} ];\\
\\
(* perform explicit sum over AL indices (expand summation convention) *)\\
fetampl =  FetSumFactor[ fetampl, AL, \{1, 3\} ];\\
\\
(* Neutralino mass matrix MN and MHMN = MN{\textasciicircum}+MN *)\\
MN = Table[0,\{i,1,4\},\{j,1,4\}];\\
MN[[1,1]] = M1;\\
MN[[2,2]] = M2;\\
MN[[3,4]] = MN[[4,3]] = mu;\\
MN[[1,3]] = MN[[3,1]] = - e v1/2/cw;\\
MN[[1,4]] = MN[[4,1]] =   e v2/2/cw;\\
MN[[2,3]] = MN[[3,2]] =   e v1/2/sw;\\
MN[[2,4]] = MN[[4,2]] = - e v2/2/sw;\\
\\
MHMN = ConjugateTranspose[MN].MN;\\
\\
(* Slepton mass matrix ML2 (neglecting small Yukawa contributions) *)\\
ML2 = Table[0,\{i,1,6\},\{j,1,6\}];\\
\\
For[i=1,i<4,i++, For[j=1,j<4,j++,\\
    ML2[[i,j]] = e{\textasciicircum}2 (v1{\textasciicircum}2-v2{\textasciicircum}2) (1-2 cw{\textasciicircum}2)/(8 sw{\textasciicircum}2 cw{\textasciicircum}2) KroneckerDelta[i,j] + MLL[j,i];\\
    ML2[[i+3,j+3]] = - e{\textasciicircum}2 (v1{\textasciicircum}2-v2{\textasciicircum}2) /(4 cw{\textasciicircum}2) KroneckerDelta[i,j] + MRR[i,j];\\
    ML2[[i,j+3]] = v1/Sqrt[2] AL[i,j];\\
    ML2[[i+3,j]] = Conjugate[ML2[[i,j+3]]];\\
  ] ];\\
\\
(* slepton MI matrix *)\\
ML2I = ML2;\\
For[i=1,i<7,i++, ML2I[[i,i]]=0 ]; \\
\\
(* choose indices for h->tau mu decay *)\\
fetampl = fetampl /. A->2 /. B->3 /. K->2;\\
\\
(* substitute real form of neutralino and slepton mass matrices *)\\
For[i=1,i<5,i++,
  For[j=1,j<5,j++,\\
    fetampl = fetampl /. MF[N,i,j] -> MN[[i,j]] /. MHM[N,i,j] -> MHMN[[i,j]];\\
  ];
];\\
\\
For[i=1,i<7,i++,
  For[j=1,j<7,j++,\\
    fetampl = fetampl /. MS2[L,i,j] -> ML2[[i,j]]  /. MS2I [L,i,j] -> ML2I[[i,j]];\\
  ];
];\\
\\
(* remove unnecessary c.c. from real parameters *)\\
fetampl = fetampl // FunctionExpand;\\
fetampl = fetampl /. Conjugate[e] -> e /. Conjugate[cw] -> cw \\
\hspace*{35mm} /. Conjugate[sw] -> sw /. Conjugate[v1] -> v1 \\
\hspace*{35mm} /. Conjugate[v2] -> v2 // Expand;\\
(* for simplicity assume real M1 *)\\
fetampl = fetampl /. Conjugate[M1] -> M1;\\[3mm]
}
At this stage the flavor violating terms could come from the FET
expansion or directly from the trilinear $A_l$ terms explicitly
present in the Higgs-slepton vertex of~\eq{eq:hll}, so their product
could be of a order higher then 1.  In addition, leaving the ${\cal
O}(M_W^2/M_{SUSY}^2)$ terms in the arguments of loop functions is
inconsistent, as similar terms have been neglected in expanding
neutralino mass matrices to 0th order only. Thus, further
simplifications should be done:\\[3mm]
{\small \tt
(* kill higher order MI terms *)\\
fetampl = fetampl /. AL[A\_,B\_] -> eps AL[A,B] /. ALP[A\_,B\_] -> eps ALP[A,B] /. \\
                     MLL[A\_,B\_] -> eps MLL[A,B] /. MRR[A\_,B\_] -> eps MRR[A,B];\\
fetampl = fetampl /. AL[A\_,A\_] -> AL[A,A]/eps /. ALP[A\_,A\_] -> ALP[A,A]/eps /. \\
                     MLL[A\_,A\_] -> MLL[A,A]/eps /. MRR[A\_,A\_] -> MRR[A,A]/eps;\\
fetampl = Normal[Series[fetampl,\{eps,0,1\}]] /. eps->1;\\
\\
(* neglect terms O(MW{\textasciicircum}2/MSUSY{\textasciicircum}2) in loop function arguments *)\\
fetampl = Simplify[fetampl]
/. e{\textasciicircum}2(v1{\textasciicircum}2 + v2{\textasciicircum}2)
-> 0 \\
\hspace*{35mm} /. e{\textasciicircum}2(v1{\textasciicircum}2 -
v2{\textasciicircum}2) -> 0
/. e{\textasciicircum}2(-v1{\textasciicircum}2 +
v2{\textasciicircum}2) -> 0;\\
\\
(* define and substitute explicit Higgs mixing matrix and vev's *)\\
ZH = Table[0,\{i,1,2\},\{j,1,2\}];\\
ZH[[1,1]] = ZH[[2,2]] = Cos[alpha]; \\
ZH[[1,2]] = - Sin[alpha];\\
ZH[[2,1]] = Sin[alpha];\\
v1 = 2 MW sw/e Cos[beta];\\
v2 = 2 MW sw/e Sin[beta];\\
For[i=1,i<3,i++,
  For[j=1,j<3,j++,
    fetampl = fetampl /. ZR[i,j]->ZH[[i,j]]
  ]
];\\[3mm]
}
In spite of lengthy and complicated intermediate expressions,
difficult to obtain without the use of computer, the final result is
compact and simple:\\[3mm]
{\small \tt
fetampl = 3 e$^2$ M1/(2 Sqrt[2] cw$^4$) (\\
+ 2/3 cw$^2$ AL[2,2] c0[1,{MLL[3,3],MLL[2,2]},MRR[2,2],M1$^2$] MLL[2,3]
Sin[alpha]\\
+ 2/3 cw$^2$ AL[3,3] c0[1,MLL[3,3],{MRR[3,3],MRR[2,2]},M1$^2$] MRR[3,2]
Sin[alpha]\\
+ (2 cw$^2$ c0[0,MLL[3,3],MRR[2,2],M1$^2$] Sin[alpha]\\
+ 2 MW$^2$ ((4 cw$^2$ - 3) c0[1,MLL[3,3],{MLL[3,3],MRR[2,2]},M1$^2$]\\
+ 2 sw$^2$ c0sq[1,{MLL[3,3],MRR[2,2]},M1$^2$]) Cos[beta]
Sin[alpha+beta])/3 AL[3,2] )\\[3mm]
}
As expected, it is linear in the three flavor violating slepton mass
insertions, {\tt MLL[2,3], MRR[2,3]} and {\tt AL[3,2]}, with the
coefficients given in a simple analytical form.

The divided differences of loop functions can be further expanded with
the use of {\tt FetExpandDividedDifferences} routine. The latter step
may be in particular required if the diagonal slepton mass terms are
degenerated and divided differences should be replaced by relevant
derivatives.  This may require some care: e.g. the $ c0sq$ function
has been earlier defined in the code as $\tt c0sq[m1^2,m2^2]\equiv
c0[m1^2,m1^2,m2^2]$, so the divided differences expansion should be
done reverting the redefinition:\\[3mm]
{\small \tt
FetExpandDividedDifferences[c0sq[ 1, \{MLL[3,3],MRR[2,2]\},M1$^2$], \\
\hspace*{7cm} c0sq[a\_,b\_] -> c0[a,a,b] ] =\\[2mm]
(c0[MLL[3,3],MLL[3,3],M1$^2$] -
c0[MRR[2,2],MRR[2,2],M1$^2$])/(MLL[3,3] - MRR[2,2])\\[3mm]
}
Obviously, the same program could be easily adapted to expand other
diagrams for this process and finally to calculate the decay branching
ratio directly in terms of parameters of the initial MSSM Lagrangian.
If necessary, with minimal modifications, it could be also used to
obtain higher order terms, both in flavor violating mass insertions
and in $M_W^2/M_{SUSY}^2$ powers. Such higher order terms may appear
important, especially when lower order MI powers cancel out, as it
often happens for rare decays in the MSSM - see e.g. discussion of
$t\to ch,uh$ decays in Ref.~\cite{Dedes:2014asa}, where the accuracy
of the results expanded by FET technique has been compared vs. the
exact numerical calculations performed in mass eigenstates basis with
the use of \code{} library~\cite{Rosiek:2010ug, Crivellin:2012jv,
  Rosiek:2014sia}.

\section{Summary}
\setcounter{equation}{0}
\label{sec:summary}

We presented \MMI{} v1.0, a Mathematica package for an automatic
expansion of transition amplitudes calculated in the mass eigenstates
basis in terms of series of mass insertions. The expressions for the
mass eigenstates amplitudes are usually more easier to calculate
diagrammatically and better suited for the numerical computations but
depend on the initial (interaction basis) Lagrangian parameters in a
complicated way.  The \MMI{} routines allow to obtain an analytical
approximation for the transition amplitudes directly as a power series
in terms of the off-diagonal entries of mass matrices, without the
need of a separate calculation of the Feynman diagrams with mass
insertions as vertices.

The package is general enough to expand any amplitude, of any loop
order, in any model involving scalar, fermion (Dirac or Majorana) or
vector particles, expressing the result in terms of the divided
differences of the loop functions.  It can perform the Mass Insertion
expansion to high orders, limited only by the combinatorial
complication of the result and growing computational time.

In addition, \MMI{} provides several auxiliary functions allowing to
express divided differences of any order in terms of the initial
function used to generate them, and to manipulate and expand
expressions containing factors with repeating indices, assumed to be
implicitly summed over.

The current version of \MMI{} Mathematica code and its manual can be
downloaded from the address
\begin{center}
\webpage
\end{center}

\section*{Acknowledgments}

This work was supported in part by Polish National Science Centre
under research grants DEC-2012/05/B/ST2/02597 and
DEC-2014/15/B/ST2/02157.  The author would like to thank University of
Ioannina and CERN for the hospitality during his stays there. I would
also like to express my gratitude to Michalis Paraskevas for thorough
testing the code and proofreading the manuscript.

\newpage


\bibliographystyle{elsarticle-num}
\bibliography{mmi}{}

\end{document}